\documentclass[preprint,amsmath,amssymb,showpacs]{revtex4}
\usepackage{epsfig}
\usepackage{dcolumn}
\usepackage{graphicx}
\usepackage{subfigure}

\begin{document}

\title{Theory of exciton fine structure in semiconductor quantum dots: quantum dot anisotropy and lateral electric field}
 \author{Eugene Kadantsev and Pawel Hawrylak}
\affiliation{Quantum Theory Group, Institute for Microstructural Sciences,
National Research Council, Ottawa, Canada K1A 0R6}
\email{ekadants@babylon.phy.nrc.ca}
\date{\today}

\begin{abstract}
Theory of exciton fine structure in semiconductor quantum dots and its dependence on quantum dot anisotropy 
and external lateral electric field is presented. The effective exciton Hamiltonian including  long range electron-hole exchange  interaction is 
derived within the $k*p$ effective mass approximation (EMA). 
The exchange matrix elements of the Hamiltonian are expressed explicitly in terms of electron and hole 
envelope functions. The matrix element responsible for the ``bright" exciton splitting 
is identified and analyzed. An excitonic fine structure for a model quantum dot with quasi- 
two-dimensional anisotropic harmonic oscillator (2DLAHO) confining potential is analyzed as a function of the shape 
anisotropy, size and applied lateral electric field.  
\end{abstract}
\maketitle

\section{Introduction}

One of the promising  applications~\cite{Benson2000,Dalacu2009} of semiconductor 
quantum dots (QDs)~\cite{Hawrylak1997, Hawrylak2003} for  quantum cryptography is the generation of 
entangled photon pairs (EPPs) 
on demand~\cite{Shields2007, Stevenson2006, Akopian2006, Greilich2006}. 
EPPs can be generated with great efficiency via the biexciton cascade 
process (BCP)~\cite{Benson2000} in which a biexciton 
radiatively decays into the ground state via two different indistinguishable paths involving 
two intermediate dipole-active (``bright") exciton states. The major 
barrier to EPPs generation in BCP is the splitting of the 
intermediate ``bright" exciton levels which distinguishes the paths of radiative 
decay and, as a result, destroys the entanglement. The  splitting  and mixing of the two bright exciton states 
is controlled by the long ranged electron-hole exchange (LRE) interaction and depends on dot asymmetry 
and applied magnetic field~\cite{Bayer2002,Stevenson2006,Greilich2006}.
Lateral electric fields have been applied in the hope to control the dot anisotropy and hence anisotropic 
exchange splitting\cite{kowalik2005,vogel2007,Reimer2008}.

Better understanding of the electron-hole exchange interaction in QDs, particularly its LRE part, 
should help in the development of EPP generation schemes. 

The exchange integral  can be decomposed in real space into long-range 
part, i.e., exchange interaction between two ``transition densities" localized in 
two different Wigner-Seitz (WS) cells and short-range part, i.e. exchange interaction within 
a single WS cell. A closely related decomposition of exchange into analytical and 
non-analytical part exists in the reciprocal space. In what follows, we will 
 use the ``real space" definition of long-range and short-range exchange (SRE). We will 
reserve the use of words ``local exchange" and ``nonlocal exchange" to describe the type of 
integrals which contribute to the exchange matrix elements. 

LRE electron-hole interaction in bulk semiconductors was investigated almost 
40 years ago~\cite{Knox1963, Denisov1973, Bir1972} and it is now well established that 
LRE splits the energy  levels of  ``bright" excitons. For example, in zinc-blende direct 
band gap semiconductors with four-fold degenerate valence band ${\rm \Gamma_{8v}}$ and two-fold
degenerate conduction band ${\rm \Gamma_{6c}}$, the ground state exciton is eight-fold degenerate. 
The addition of SRE interaction into the excitonic 
Hamiltonian will split the eight-fold degenerate ground state into ``dark" and ``bright" multiplets
with degeneracies of five and three. The addition of LRE interaction will further modify 
the fine structure by splitting three-fold degenerate ``bright" 
exciton level into two transverse excitons with $J_z = \pm 1$ and one
longitudinal exciton with $J_z = 0$ . 

Recent advances in single QD spectroscopies motivated re-examination of electron-hole exchange in
systems with reduced dimensionality within the framework of envelope function approximation~\cite{Takagahara1993,Sham1993,Efros1996,Gupalov1998,Takagahara2000,Gupalov2000,Glazov2007} 
and, more recently, from the point of view of empirical atomistic 
tight-binding approximation~\cite{Franceschetti1998,Goupalov2001,Bester2003,He2008}. 

Within the envelope function approximation, Takagahara~\cite{Takagahara1993} have 
shown that if the electron-hole pair envelope contains only $Y_{00}(\theta,\phi)$ 
angular momentum component, then the long-range part of the electron-hole exchange vanishes. 
In Refs.~\cite{Sham1993,Gupalov1998}, fine structure of localized exciton levels in quantum wells 
was considered. Efros~{\it el. al.}~\cite{Efros1996} and Gupalov~{\it et. al.}~\cite{Gupalov2000} 
investigated band-edge excitonic fine structure of spherical {\rm CdSe} nanocrystals. 
In Ref.~\cite{Takagahara2000}, Takahagara derived effective 
eight-band excitonic Hamiltonian which takes into account electron-hole exchange 
interaction. In particular,  within the envelope function 
formalism of Refs.~\cite{Takagahara1993,Takagahara2000}, LRE is a dipole-dipole interaction. 
Takagahara~\cite{Takagahara2000} applied his scheme to investigate excitonic 
fine structure of disk-like {\rm GaAs/AlGaAs} QDs as a function of QD anisotropy and size. 
It was  demonstrated numerically in Ref.~\cite{Takagahara2000}
that LRE vanishes for the ground state ``bright" exciton doublet 
in QDs with  rotationally symmetric 3D confining  potential. 
A similar to that of Ref.~\cite{Takagahara2000} representation of the electron-hole exchange 
was discussed by Maialle~{\it et. al.}~\cite{Sham1993} in connection with the exciton spin 
dynamics in quantum wells.

Within atomistic empirical tight-binding approach it has been 
demonstrated that the physical origin of long-range exchange interaction 
might be different from that of bulk~\cite{Franceschetti1998} 
and that LRE has a nonvanishing magnitude even in ``shape-symmetric" dots~\cite{Bester2003,He2008}. 
The former and the latter  stem from the loss of local orthogonality on the unit cell scale 
between electron and hole single-particle orbitals compared to the bulk case and ``reduced" 
symmetry of the atomistic confining potential, respectively. 
For ${\Gamma_6 \times \Gamma_7}$ exciton, Gupalov {\it et. al.}~\cite{Goupalov2001}  
identified the dipole-dipole and monopole-monopole contributions to the LRE with 
the intra-atomic and interatomic transition densities, respectively. 

In this work, we will re-examine electron-hole exchange within the framework of envelope 
approximation. We note that Takagahara's condition~\cite{Takagahara1993} 
for the vanishing of LRE in spherically symmetric quantum dots
constitutes only a  {\it sufficient} condition for quenching of LRE. 
It does not explain, for example, why LRE vanishes in disk-like QDs with 
rotational ($C_{\infty v}$) symmetry. These QDs are ``squeezed" in one direction 
and the ground state envelopes of single-particle states will contain angular momentum 
components that are higher than $Y_{00}(\theta,\phi)$. Motivated by recent experiments,
we will  investigate, within our model,  the effects of the external electric field and size-scaling of LRE interaction.  

We derive here effective ``four band" excitonic 
Hamiltonian which includes effects of LRE interaction. 
The elements of the effective Hamiltonian are expressed 
explicitly in terms of envelope functions. The ``microscopic parts" of single-particle 
orbitals are integrated out and enter the effective Hamiltonian as numerical parameters. 
The number of bands  is  truncated to two conduction and 
two valence bands for the simplicity of the interpretation. 
We present explicit expression  for the matrix element 
responsible for the ``bright" exciton splitting and, therefore, establish a 
new sufficient condition for LRE quenching. 

An excitonic fine structure for a model system in which the confining potential 
has a form of  two-dimensional-like anisotropic harmonic oscillator 
(2DLAHO) is considered  as a function of lateral anisotropy. 
We present an explicit expression 
for the ``bright" splitting of excitonic ground state as a function of 
lateral anisotropy. It is found, in agreement with 
previous work~\cite{Takagahara2000}, that, within the envelope approximation,  
the ``bright" ground state exciton splitting vanishes in the case of laterally isotropic 
confining potential. 
The quenching of ``bright" exciton splitting coincides with vanishing of the matrix element 
responsible for the ``bright" exciton splitting in our effective Hamiltonian.

We also analyze the effect of the lateral electric field 
on the excitonic fine structure of our model quantum dot. We find that the ``bright" exciton
splitting decreases due to the spatial separation of electron and hole envelopes. 

Finally, the scaling of the ``bright" exciton splittings with system's size is analyzed. It is 
found that the scaling of ``bright" exciton splittings differ from the laws established 
using simple dimensionality arguments. 

Effective units of length and energy are used throughout unless 
otherwise specified. The lengths are measured in effective Bohrs 
$a_{eB} = \epsilon \hbar^2 / (m^* e^2)$, where $\epsilon$ is dielectric constant and
$m^*$ is the conduction band effective mass. Energies are measured in effective 
Hartrees,  1~{\rm H}~$ = m^* e^4 / (\epsilon \hbar)^2 = 2$~{\rm Ry}$^*$. 
For example, using material parameters for {\rm GaAs}, $a_{eB} = 97.9$~\AA~and 
1~{\rm H}~=~11.86 meV. 
 
\section{Theory of exciton fine structure in envelope function approximation} 
We now describe the single particle states and the exciton fine structure in the envelope function approximation.
The single-particle orbitals of the electron in a quantum dot are two-component spinors 
(two-row columns)  written as
\begin{equation}
  \phi(\mathbf{r}) =  
 \left( 
\begin{array}{c} 
 \phi_a(\mathbf{r})   \\
 \phi_b(\mathbf{r})   
\end{array}
\right) = \phi_a(\mathbf{r}) | \alpha \rangle + \phi_b(\mathbf{r}) | \beta \rangle, 
\quad |\alpha \rangle=  \left( 
\begin{array}{c} 
 1   \\
 0
\end{array}
\right), 
|\beta \rangle =  \left( 
\begin{array}{c} 
 0   \\
 1
\end{array}
\right). 
\end{equation} 
A dagger sign $^{\dagger}$  will denote complex conjugation 
for complex quantities. 

With $c^{\dagger}$ and $h^{\dagger}$ ($c$ and $h$)  electron and hole 
creation (annihilation) operators,
  the interacting electron-hole Hamiltonian can be written as
\begin{eqnarray}
\label{HAM}
   \hat{H}_X & = & \hat{H}_e + \hat{H}_h + \hat{H}_{int}, \quad
   \hat{H}_e  =  \sum_{i} \varepsilon_{i}^e c_i^{\dagger} c_i, \quad
   \hat{H}_h  =  \sum_{j} \varepsilon_{j}^h h_j^{\dagger} h_j, \nonumber \\
   \hat{H}_{int} & = &    
\sum_{ijkl} c^{\dagger}_i h^{\dagger}_j h_k c_l \left (-\frac{1}{\epsilon}V_{ikjl}^C+ V_{iklj}^E \right),
\nonumber \\
&& V_{ijkl} = V(i,j;k,l) = \int \int d \mathbf{r}_1 d \mathbf{r}_2 
\frac{\phi_i^{\dagger}(\mathbf{r}_1)\phi_j^{\dagger}(\mathbf{r}_2) \phi_k(\mathbf{r}_2) \phi_l(\mathbf{r}_1) }{|\mathbf{r}_1-\mathbf{r}_2|}
\end{eqnarray} 
where $\hat{H}_e$, $\hat{H}_h$, and $\hat{H}_{int}$
are electron, hole, and electron-hole interaction Hamiltonian, 
respectively. The electron-hole interaction Hamiltonian consists of two parts: 
$V^C/\epsilon$ is the direct electron-hole Coulomb attraction screened by static dielectric 
constant $\epsilon$, $V^E$  is the electron-hole exchange interaction. 
The question of screening of electron-hole exchange 
is a  subtle one and it is generally agreed that, at least, parts of the electron-hole exchange should 
be screened. In Ref.~\cite{Ivchenko2004} (page 252), the long-range electron-hole exchange interaction 
in reciprocal space is screened by the high-frequency dielectric constant. We will assume, for now, that 
our exchange interaction $V^E$ contains screening implicitly. 

To obtain excitonic states, Hamiltonian~(\ref{HAM}) is diagonalized 
in the basis of all electron-hole pairs of the type 
$ c^{\dagger}  h^{\dagger} | g.s. \rangle$, 
where $|g.s. \rangle$ denotes a many-body state with fully occupied valence and 
empty conduction bands. 

In this work, the hole and electron single-particle  states are computed  
in the effective mass approximation (EMA)~\cite{Wojs1996,Ivchenko2004} which neglects band coupling
in the single-particle states.
In EMA, the hole $\phi_{h}$ and electron $\phi_{e}$ single-particle states are uniquely 
specified by band label (valence $v j_z'$ and conduction $c \sigma'$) and envelop index ($r$ and $p$), 
for example, 
\begin{eqnarray}
\phi_{h}(\mathbf{r}) & = &  F_{v j_z'}^{r}(\mathbf{r}) u_{v j_z'}(\mathbf{r}), \nonumber \\
\phi_{e}(\mathbf{r}) & = &  F_{c \sigma'}^{p}(\mathbf{r}) u_{c \sigma'}(\mathbf{r}) . \nonumber \\ 
\end{eqnarray}
Here, $u(\mathbf{r})$ is the periodic part of Bloch eigenstate at $\mathbf{k}=0$. 
In what follows, $u(\mathbf{r})$ is a two-component spinor. We will assume that ``bra" 
electron-hole pair is described by indices $c \sigma s$ and $v j_z  q$,
respectively. 

The excitonic Hamiltonian matrix element in the basis of electron-hole pairs is
\begin{eqnarray}
&& \langle g.s. | h_{v j_z q}  c_{c \sigma s} \hat{H}_X c^{\dagger}_{c \sigma' p}  h^{\dagger}_{v j_z' r} | g.s. \rangle
= \delta_{v j_z q,v j_z' r} \delta_{c \sigma s,c \sigma' p} ( \varepsilon_{v j_z q}^h+ \varepsilon_{c \sigma s}^e) 
\nonumber \\
&& - V^C(c \sigma s,v j_z' r;v j_z q , c \sigma' p)/\epsilon + V^E(c \sigma s, v j_z' r; c \sigma' p, v j_z q)
\end{eqnarray}

The effective Hamiltonian which involves only envelope functions is obtained by integrating 
out ``microscopic" degrees of freedom  ($u$).  
The derivation follows  that of Ref.~\cite{Takagahara2000}. 
Assuming $T_d$  symmetry of the crystal, two conduction band 
``microscopic" functions  $|c 1/2 \rangle$ and $|c -1/2 \rangle$ 
with spin projections $s_z = 1/2$ and $s_z = -1/2$ are written  as
\begin{equation}
\label{EL1}
 u_{c 1/2}(\mathbf{r}) = \langle \mathbf{r} |c 1/2 \rangle = \chi_s (r) 
 \left( 
\begin{array}{c} 
 Y_{00}(\hat{r})   \\
 0   
\end{array}
\right), \quad
u_{c -1/2}(\mathbf{r}) =  \langle \mathbf{r} |c -1/2 \rangle = 
\chi_s (r) 
 \left( 
\begin{array}{c} 
 0    \\
 Y_{00}(\hat{r})   
\end{array}
\right).  
\end{equation} 
In Eq.~(\ref{EL1}), we assumed that ``microscopic" conduction band function are of pure 
{\it s} symmetry. 

Two valence band ``microscopic" functions
$|v-3/2\rangle$ and $|v+3/2\rangle$  with hole angular momentum projections 
$j_z = -3/2$ and $j_z = 3/2$ are taken as
\begin{eqnarray}
\label{HOLES1}
&& u_{v-3/2}(\mathbf{r}) = \langle \mathbf{r} |v-3/2\rangle = \chi_p (r) 
 \left( 
\begin{array}{c} 
 Y_{11}(\hat{r})   \\
 0   
\end{array}
\right), \quad 
u_{v+3/2}(\mathbf{r}) = \langle \mathbf{r} |v+3/2\rangle = 
\chi_p (r) 
 \left( 
\begin{array}{c} 
 0   \\
 Y_{1-1}(\hat{r})
\end{array}
\right). \nonumber \\
\end{eqnarray} 
In Eq.~(\ref{HOLES1}) we assumed that ``microscopic" valence band functions are of pure 
{\it p} symmetry. The eigenfunctions of angular momentum $\chi_p (r)  Y_{11}(\hat{r})$ and $\chi_p (r) Y_{1-1}(\hat{r})$ 
can be expressed in terms of Cartesian functions $p_x$ and $p_y$. We, therefore, 
are neglecting $p_z$ contribution to the ``microscopic" valence band functions. 

In what follows, we give a brief derivation of the effective excitonic Hamiltonian.

\subsection{Calculation of Coulomb Direct Matrix Elements.}

The unscreened Coulomb direct matrix element is given by 
\begin{eqnarray} 
&& V^{C}(c \sigma s,v j_z' r;v j_z q , c \sigma' p)
=  \int \int d\mathbf{r}_1 d\mathbf{r}_2 
\frac{ q_1(\mathbf{r}_1) q_2(\mathbf{r}_2)}{|\mathbf{r}_1 - \mathbf{r}_2|} \nonumber \\
&& q_1(\mathbf{r}_1) = F_{c \sigma}^{s \dagger}(\mathbf{r}_1) u_{c \sigma}^{\dagger}(\mathbf{r}_1)
F_{c \sigma'}^{p}(\mathbf{r}_1) u_{c \sigma'}(\mathbf{r}_1),  \nonumber \\
&& q_2(\mathbf{r}_2) = F_{v j_z'}^{r \dagger}(\mathbf{r}_2) u_{v j_z'}^{\dagger}(\mathbf{r}_2)
F_{v j_z}^{q}(\mathbf{r}_2) u_{v j_z}(\mathbf{r}_2),  \nonumber \\
\end{eqnarray} 
where we explicitly presented electron and hole single-particle orbitals as a product of 
an envelope $F$ and a ``microscopic" part $u$. The unscreened Coulomb attraction matrix element is just Coulomb interaction between 
two ``transition densities", where each ``transition density" is a product of  two
electron orbitals or 
two valence orbitals. The matrix element is approximated as
\begin{equation}
V^{C}(c \sigma s,v j_z' r;v j_z q , c \sigma' p) =  
\delta_{c \sigma c \sigma'} \delta_{v j_z v j_z'} \int \int d\mathbf{r}_1 d\mathbf{r}_2
\frac{F_{c \sigma}^{s \dagger}(\mathbf{r}_1) F_{c \sigma'}^{p}(\mathbf{r}_1) 
F_{v \sigma'}^{r \dagger}(\mathbf{r}_2) F_{v \sigma}^{q}(\mathbf{r}_2)}{|\mathbf{r}_1-\mathbf{r}_2|} .
\end{equation} 

\subsection{Calculation of Coulomb Exchange Matrix Elements.}

The exchange matrix element can be written as
\begin{eqnarray} 
\label{EXINT1}
&&  V^{E}(c \sigma s,v j_z' r; c \sigma' p, v j_z q) = \int \int d\mathbf{r}_1 d\mathbf{r}_2 
\frac{ q_1(\mathbf{r_1})  q_2 (\mathbf{r_2})  }{|\mathbf{r}_1 - \mathbf{r}_2|}, \nonumber \\
&& q_1(\mathbf{r}_1) = F_{c \sigma}^{s \dagger}(\mathbf{r}_1) u_{c \sigma}^{\dagger}(\mathbf{r}_1)
F_{v j_z}^{q}(\mathbf{r}_1) u_{v j_z}(\mathbf{r}_1), \nonumber \\
&& q_2(\mathbf{r}_2) = F_{c \sigma'}^{p}(\mathbf{r}_2) u_{c \sigma'}(\mathbf{r}_2)
F_{v j_z'}^{r \dagger}(\mathbf{r}_2) u_{v j_z'}^{\dagger}(\mathbf{r}_2).  \nonumber \\
\end{eqnarray} 
The exchange matrix element can be thought of as a Coulomb interaction of two 
``transition densities", where each ``transition density" is a product of electron-hole 
single-particle orbitals. 

We decompose exchange integral~(\ref{EXINT1}) into the short-range  and 
long-range  contributions in real space. We will refer to the whole integration region  as 
Born-von Karmen cell ({\rm BvK} cell) and to the individual unit cell within {\rm BvK} cell as 
Wigner-Seitz cell ({\rm WS} cell). 
The double integration over {\rm BvK} cell (which consists of $N_{cell}$ {\rm WS} cells) 
is replaced by $N_{cell} \times N_{cell}$ integrals over {\rm WS} cells
\begin{equation}
\label{DECOMP}
\int_{\mathbf{r}_1 \in BvK} \int_{\mathbf{r}_2 \in BvK} d \mathbf{r}_1 d \mathbf{r}_2 \rightarrow 
\sum_{i=1}^{N_{cell}} \int_{\mathbf{r}_1 \in WS(\mathbf{R}_i)} 
\int_{\mathbf{r}_2 \in WS(\mathbf{R}_i)} d \mathbf{r}_1 d \mathbf{r}_2
+ \sum_{{\tiny \begin{array}{c} 
 i,j=1   \\
 i \neq j
\end{array}}}^{N_{cell}} \int_{\mathbf{r}_1 \in WS(\mathbf{R}_i)} 
\int_{\mathbf{r}_2 \in WS(\mathbf{R}_j)} d \mathbf{r}_1 d \mathbf{r}_2, 
\end{equation} 
where $\mathbf{R}_i, \mathbf{R}_j$  label positions of the {\rm WS} cells. 
The first term in (\ref{DECOMP}) consists  of $N_{cell}$ integrals in which $\mathbf{r}_1$ and $\mathbf{r}_2$
go over the same cell, whereas the second term consists of  $N_{cell} \times (N_{cell}-1)$ integrals in which 
$\mathbf{r}_1$ and $\mathbf{r}_2$ go over two distinct {\rm WS} cells. The first sum in 
which $\mathbf{r}_1$ and $\mathbf{r}_2$ go over the same cell
is the short-range exchange $V^{E}_{SR}$, whereas the second sum in which 
$\mathbf{r}_1$ and $\mathbf{r}_2$ go over two different {\rm WS} cells is the
long-range exchange $V^{E}_{LR}$.

After some algebra which involves multipole expansion of $1/|\mathbf{r}_1 - \mathbf{r}_2|$, we 
obtain the following expression for the exchange matrix element
\begin{eqnarray}
&& V^E(c \sigma s, v j_z' r ;c \sigma' p,v j_z q) = 
V^{E}_{SR}(c \sigma s, v j_z' r ;c \sigma' p,v j_z q)  + 
V^{E}_{LR}(c \sigma s, v j_z' r ;c \sigma' p,v j_z q), \nonumber \\
&& V^{E}_{SR}(c \sigma s, v j_z' r ;c \sigma' p,v j_z q) = 
E_{SR} (H_{SR}^{int})^{c\sigma,v j_z}_{c \sigma',v j_z'}
\int d \mathbf{r} F_{c \sigma}^{s \dagger}(\mathbf{r}) 
F_{v j_z}^{ q} (\mathbf{r}) 
F_{c \sigma'}^{ p}(\mathbf{r}) F_{v j_z'}^{ r\dagger}(\mathbf{r}),
\nonumber \\
&& V^{E}_{LR}(c \sigma s,v j_z' r;c \sigma' p,v j_z q)  = 
-\frac{4 \pi}{3} \mu^2 \left( (\mathbf{d}_{c\sigma v j_z }^0)^{\dagger} \cdot \mathbf{d}_{c\sigma' v j_z' }^0 \right)
\int  d \mathbf{r} F_{c \sigma}^{s \dagger}(\mathbf{r}) F_{v j_z}^{q}(\mathbf{r})
F_{c \sigma'}^{p}(\mathbf{r}) F_{v j_z'}^{r \dagger}(\mathbf{r}) \nonumber \\
&& - \mu^2 \sum_{\gamma,\delta=1}^3  (\mathbf{d}_{c \sigma v j_z }^0)^{\dagger}_{\gamma}  
(\mathbf{d}_{c\sigma' v j_z' }^0)_{\delta}  
\int \int \left(\frac{\partial^2 F_{c \sigma}^{s \dagger}(\mathbf{r}_1) F_{v j_z}^{q}(\mathbf{r}_1)}{ \partial r_1^{\gamma} \partial r_1^{\delta}} 
\right)
\frac{F_{c \sigma'}^{p}(\mathbf{r}_2) F_{v j_z'}^{r \dagger}(\mathbf{r}_2)}{|\mathbf{r}_1-\mathbf{r}_2|}
 d \mathbf{r}_1 d \mathbf{r}_2 \nonumber \\ 
\end{eqnarray} 
where $E_{SR}$ and $\mu^2$ are two numerical constants parameterizing short- and long-range 
exchange interaction, respectively, $(H_{SR}^{int})^{c\sigma,v j_z}_{c \sigma',v j_z'}$ 
is an element of short-range exchange ``microscopic" matrix
\begin{equation}
\label{HSRINT}
\mathbf{H_{SR}^{int}} = 
 \left( 
\begin{array}{cccccccccc} 
 &  &|c \sigma', v j_z'\rangle  & 
    |c \sigma', v j_z'\rangle  & 
    |c \sigma', v j_z'\rangle  & 
    |c \sigma', v j_z'\rangle   \\
 &  &|\frac{1}{2},-\frac{3}{2} \rangle &  |\frac{1}{2} ,\frac{3}{2} \rangle 
   &|-\frac{1}{2},-\frac{3}{2} \rangle & |-\frac{1}{2}, +\frac{3}{2} \rangle \\
|c \sigma, v j_z \rangle & | \frac{1}{2}, -\frac{3}{2} \rangle & 1  & 0 & 0  & 0 \\       
|c \sigma, v j_z \rangle & | \frac{1}{2}, +\frac{3}{2} \rangle & 0  & 0 & 0  & 0 \\   
|c \sigma, v j_z \rangle & | -\frac{1}{2},  -\frac{3}{2} \rangle & 0  & 0 & 0  & 0 \\   
|c \sigma, v j_z \rangle & | -\frac{1}{2},  +\frac{3}{2} \rangle & 0 &  0 & 0  & 1 \\ 
\end{array}
\right) ,  
\end{equation} 
and $\mathbf{d}_{c\sigma v j_z }^0$ are ``microscopic" dipole elements
\begin{equation}
\label{MICDIPOLES}
\begin{array}{ccccc} 
               &  & (\mathbf{d}_{c\sigma vj_z }^0)_x 
                  & (\mathbf{d}_{c\sigma vj_z }^0)_y 
	          & (\mathbf{d}_{c\sigma vj_z }^0)_z \\
|c \sigma,  v j_z \rangle & |\frac{1}{2},  -\frac{3}{2} \rangle &  -1  & -i  & 0 \\
|c \sigma,  v j_z \rangle &|\frac{1}{2},  +\frac{3}{2} \rangle &   0 & 0 & 0 \\
|c \sigma,  v j_z \rangle &|-\frac{1}{2},   -\frac{3}{2} \rangle &    0 & 0 & 0 \\ 
|c \sigma,  v j_z \rangle &|-\frac{1}{2},   +\frac{3}{2} \rangle & 1 & -i & 0  . 
\end{array} 
\end{equation}

The numerical parameters $E_{SR}$ and $\mu^2$ can be determined 
as to reproduce the excitonic fine structure in bulk semiconductors and, therefore, 
implicitly contain screening effects. 

Taking into account~(\ref{HSRINT}) and~(\ref{MICDIPOLES}),
a ``block" of Hamiltonian corresponding to the exchange interaction between two electron-hole 
pairs $F_c^p F_v^r$ and $F_c^s F_v^q$ can be presented in the form
\begin{equation}
\label{MASTER}
 \left( 
\begin{array}{cccccccccc} 
 &  &|c \sigma', v j_z'\rangle  & 
    |c \sigma', v j_z'\rangle  & 
    |c \sigma', v j_z'\rangle  & 
    |c \sigma', v j_z'\rangle   \\
 &  &|\frac{1}{2},-\frac{3}{2} \rangle &  |\frac{1}{2} ,+\frac{3}{2} \rangle 
   &|-\frac{1}{2}, -\frac{3}{2} \rangle & |-\frac{1}{2}, +\frac{3}{2} \rangle \\
|c \sigma, v j_z \rangle & | \frac{1}{2}, -\frac{3}{2} \rangle & \delta_{0}^{SRE,L}+\delta_{0}^{LRE,L}+\delta_{0}^{LRE,N}  & 0 & 0  & \delta_{12}^{LRE,N} \\       
|c \sigma, v j_z \rangle & | \frac{1}{2}, +\frac{3}{2} \rangle & 0  & 0 & 0  & 0 \\   
|c \sigma, v j_z \rangle & | -\frac{1}{2},  -\frac{3}{2} \rangle & 0  & 0 & 0  & 0 \\   
|c \sigma, v j_z \rangle & | -\frac{1}{2},  +\frac{3}{2} \rangle & \delta_{21}^{LRE,N} &  0 & 0  & \delta_{0}^{SRE,L}+\delta_{0}^{LRE,L}+\delta_{0}^{LRE,N} \\ 
\end{array}
\right) ,  
\end{equation} 
where we separated different contributions to the exchange based on their origin (SRE or LRE) and integral type (``local" and ``nonlocal"). 
The contributions are
\begin{eqnarray}
\label{LRE1}
&& \delta_{0}^{SRE,L} = E_{SR} \int d \mathbf{r} F_{c}^{s \dagger}(\mathbf{r}) F_{v}^{ q} (\mathbf{r}) F_{c}^{ p}(\mathbf{r}) F_{v}^{ r\dagger}(\mathbf{r}),
\nonumber \\
&& \delta_{0}^{LRE,L} = -\frac{8 \pi \mu^2}{3}  \int d \mathbf{r} F_{c}^{s \dagger}(\mathbf{r}) F_{v}^{ q} (\mathbf{r}) F_{c}^{ p}(\mathbf{r}) F_{v}^{ r\dagger}(\mathbf{r}),
\nonumber \\
&&  \delta_{0}^{LRE,N} = -\mu^2 \left( R_{xx} + R_{yy} \right) \nonumber \\
 && \delta_{12}^{LRE,N} = V^{E}_{LR}(1/2 s, 3/2 r;-1/2 p, -3/2 q) = 
\mu^2 \left( R_{xx} -2i R_{xy} - R_{yy} \right), \nonumber \\
 && \delta_{21}^{LRE,N} = V^{E}_{LR}(-1/2 s,-3/2 r;1/2 p, 3/2 q) = 
\mu^2 \left( R_{xx} + 2i R_{xy} - R_{yy} \right), \nonumber \\ 
&& R_{\delta \gamma} = 
 \int \int \left\{\left( \frac{\partial^2}{\partial r_1^{\delta} \partial r_1^{\gamma}}\right)
 F_{c}^{s\dagger} F_{v}^{q} \right\} 
\frac{F_{c}^{p}(\mathbf{r}_2) F_{v}^{r\dagger}(\mathbf{r}_2)}{|\mathbf{r}_1-\mathbf{r}_2|}
 d \mathbf{r}_1 d \mathbf{r}_2, \nonumber \\
\end{eqnarray} 

With regard to the above exchange expressions we note the following:
(a)The short-range exchange causes splitting between ``bright"  
($| \frac{1}{2}, -\frac{3}{2} \rangle, |-\frac{1}{2},  +\frac{3}{2} \rangle$) 
and ``dark" doublets ($| \frac{1}{2}, +\frac{3}{2} \rangle, |-\frac{1}{2},  -\frac{3}{2} \rangle$) by 
moving ``bright" doublet up in energy by $\delta_{0}^{SRE,L}$. SRE does not split the ``bright" doublet. 
The integral describing SRE is ``local" in nature and involves overlap between two electron-hole pairs. 
(b) Long-range exchange contains expressions of two types: 
a ``local"  term $\delta_{0}^{LRE,L}$ which arises from 
$\mu^2 \left( (\mathbf{d}_{c\tau v\sigma }^0)^{\dagger} \cdot \mathbf{d}_{c\tau' v\sigma' }^0 \right)$ and 
``nonlocal" terms $\delta_{0}^{LRE,N}$, $\delta_{12}^{LRE,N}$, 
and $\delta_{21}^{LRE,N}$ which involve differentiation operators applied to 
electron-hole pair envelops. The ``nonlocal" terms describe dipole-dipole 
interaction between electron-hole transition densities. 
(c)  The splitting between ``bright"-``dark" states is affected by LRE through the ``local" LRE term $\delta_{0}^{LRE,L}$ 
and ``nonlocal" LRE term $\delta_{0}^{LRE,N}$.
LRE, in general, splits the ``bright" 
doublet by coupling electron-hole pairs with ``completely opposite" 
$z$-projections of angular momentum, for example, 
$|-\frac{1}{2},\frac{3}{2}\rangle$ and $|\frac{1}{2},-\frac{3}{2}\rangle$ . The terms which are responsible for the 
``bright" exciton splitting are $\delta_{12}^{LRE,N}$  and $\delta_{21}^{LRE,N}$ - the ``nonlocal" expressions 
arising from long-range exchange interaction. 
(d) The splitting within the ``dark" doublet is not described in our model. 

The splitting of ``bright" exciton levels vanishes provided that 
$\delta_{12}^{LRE,N}$ and $\delta_{21}^{LRE,N}$ vanish which constitutes a condition
for LRE quenching within our model. In the next section, 
we demonstrate numerically and analytically that 
this condition is fulfilled in the case of isotropic 2D HO confining 
potential. 

\section{Exciton fine structure of quasi-two-dimensional anisotropic harmonic oscillator (2DLAHO) quantum dot} 

In this section, we will investigate excitonic fine-structure for a  model quantum dot in which 
the confining potential is  of 2D-like anisotropic harmonic oscillator type. We note that
HO oscillator spectrum has been observed in self-assembled quantum dots~\cite{raymond2004}.

The EMA equations for holes are
\begin{eqnarray}
\label{HH1}
&& \hat{H}_{hh} = -\frac{1}{2 M_{||,hh}} \left(\nabla_x^2+\nabla_y^2 \right)
+\frac{1}{2} M_{||,hh} \left(\omega_{x,hh}^2 x^2 + \omega_{y,hh}^2 y^2 \right) 
 -\frac{1}{2 M_{\perp,hh}} \nabla_z^2
+\frac{1}{2} M_{\perp,hh} \omega_{z,hh}^2 z^2,  \nonumber \\
&& \omega_{x,hh} = \omega_{hh}^{0}(1+t), \quad 
   \omega_{y,hh} =  \frac{\omega_{hh}^{0}}{1+t}, \quad 
   \omega_{x,hh} \omega_{y,hh} = (\omega_{hh}^0)^2 = const, \quad  
   \omega_{z,hh} \gg \omega_{hh}^0 \nonumber, \\
\end{eqnarray} 
where $M_{||,hh}$ and $M_{\perp,hh}$ are 
effective masses, $\omega_{\gamma} (\gamma = x, y, z)$ denotes confinement frequency 
in $x$, $y$, and $z$ directions, respectively,   
$\omega^{0}_{hh}$  determines the typical energy scale of the confining potential and, together with the mass, the confinement. 
The energies and lengths are expressed in the effective units. 

EMA equations for electrons have a similar form
\begin{eqnarray}
\label{EE1}
&& \hat{H}_{ee} = -\frac{1}{2} \left(\nabla_x^2 + \nabla_y^2 + \nabla_z^2  \right)
+\frac{1}{2}  \left(\omega_{x,ee}^2 x^2 + \omega_{y,ee}^2 y^2 + \omega_{z,ee}^2 z^2\right), \nonumber \\
&& \omega_{x,ee} = \omega_{ee}^{0}(1+t), \quad 
   \omega_{y,ee} =  \frac{\omega_{ee}^{0}}{1+t}, \quad 
   \omega_{x,ee} \omega_{y,ee} = (\omega_{ee}^0)^2 = const, \quad  
   \omega_{z,ee} \gg \omega_{ee}^0
\nonumber \\   
\end{eqnarray}

The excitonic fine structure is studied as a function of anisotropy $t$. 
For example, for $t=-0.5$, $\omega_x = (1/2) \omega^0$ and $\omega_y = 2 \omega^0$, 
the confinement along $x$ is weaker than along $y$. 
For $t=0.5$, $\omega_x = (3/2) \omega^0$ and $\omega_y = 2/3 \omega^0$, 
confinement along $x$ is greater than along $y$.
In the case $t = 0$, the confining potential for holes and electrons is isotropic in the 
lateral direction.  By changing anisotropy $t$, we control the shape 
of the confining potentials for holes (\ref{HH1}) and electrons (\ref{EE1}) and, 
as a consequence, the single-particle energy spectra and eigenfunctions. 

We have assumed that the confinement in the vertical 
direction $z$  is much stronger than confinement in the lateral direction 
$\omega_{z} \gg \omega^0$. Therefore, we will always assume the ground state solution 
in $z$ direction and the single-particle energy spectra for holes (\ref{HH1}) and 
electrons (\ref{EE1}) is 2D-like
\begin{equation}
\label{2DLAHOS}
E(n,m,0) = \left(n + \frac{1}{2} \right) \omega_x + 
\left(m + \frac{1}{2} \right) \omega_y + 
\frac{1}{2} \omega_z, 
\end{equation}
where $n$ and $m$ 2DLAHO quantum numbers. In what follows, we use  numerical parameters 
$M_{||,hh}=9.930853$, $M_{\perp,hh}=5.218662$, $M_{||,ee}=1.0$, $M_{\perp,ee}=1.0$,
$\omega_{z,hh}=20.155737$, and $\omega_{z,ee}=105.185977$. 

The eigenfunctions are products 
of one-dimensional Hermite polynomials and exponential function
\begin{equation}
\psi_{nm0}(\mathbf{r}) = \psi_n(x) \psi_m(y) \psi_0(z), 
\end{equation}
where the eigenfunction in one of the Cartesian directions (for example, $x$)
is given by
\begin{eqnarray}
&& \psi_n(x) = \sqrt{\frac{1}{2^n n!}} \left(\frac{m \omega_x}{\pi} \right)^{1/4} 
\exp{\left(-\frac{m \omega_x}{2} x^2\right)} H_n(\sqrt{m \omega_x} x), \nonumber \\
&& H_n(x) = (-1)^n \exp{(x^2)} \frac{d^n}{d x^n} \exp{(-x^2)} . \nonumber \\
\end{eqnarray}
Figures~(1a),~(1b) and~(1c),~(1d) show single-particle spectra for electrons and holes, 
respectively, for two different characteristic energy scales $\omega^0$. 
Solid horizontal line  (around 52.6 and 10.1~{\rm H}  for electrons and holes, respectively) 
corresponds to the confinement energy in vertical direction $\omega_z/2$. 

The single-particle energy spectra are given by~(\ref{2DLAHOS}). The single-particle ground
state is the nodeless $s$ envelope which corresponds to $n=m=0$. The ``excited" levels are organized 
in shells referred to as $p$, $d$, $f$, and so on with characteristic ``spatial" degeneracies of 2, 3, 4, 
etc. in the case of isotropic ($t = 0$) confinement potential.
The dispersion $E(t)$ of $s$ energy level with anisotropy is weak compared to the dispersion of ``excited" envelops. 
 Strong anisotropy might lead to the level crossing where 
the single-particle energy 
of $d$-type envelope is below the single particle energy of $p$-type envelop. The level crossing
happens, for example, at $t \approx -0.3$  for the second excited state on Fig.~(1a). 
Due to the smaller effective mass of electrons, the spacing between electron single-particle 
levels is larger than that of the hole levels. 

Figures~(1e) and (1f) show ``noninteracting" electron-hole and ``interacting" exciton energies 
as a function of lateral anisotropy $t$ for two different characteristic energy scales 
$\omega^0$. The black squares on Figures~(1e) and~(1f) correspond to the ``noninteracting" 
electron-hole  pair energies 
$\delta_{v \sigma q,v \sigma' r} \delta_{c \tau s,c \tau' p} ( \varepsilon_{v \sigma q}^h+
\varepsilon_{c \tau s}^e)$.  The empty circles are obtained by diagonalizing the full excitonic 
Hamiltonian. Each of the empty circles is actually a multiplet of 4 excitons 
with fine structure determined by the electron-hole exchange interaction. The 
``noninteracting" electron-hole and  excitonic spectra look quite similar. 
The Coulomb electron-hole attraction simply decreases the exciton energy. 
The Coulomb attraction-induced mixing  of electron-hole pairs  in an exciton 
is small due to the large separation of single-particle levels on the scale of 
magnitude of screened Coulomb attraction.

In the lower part of the ``interacting" excitonic energy spectra 
and for small anisotropies $|t|  \leq 0.2$, the dispersion $E_X(t)$ strongly resembles that of the 
dispersion of single-particle levels. This happens because $\omega^0_{hh} < \omega^0_{ee}$ and
the lowest energy electron-hole pairs follow the order of hole levels $s_e s_h$, $s_e p_h$, and $s_e d_h$.  

\subsection{LRE interaction for $s$-type envelopes}

Consider the case when exciton is given 
by the product of ground state  envelopes of electron and hole single-particle states 
($s_e s_h$ exciton). In this case,
\begin{eqnarray}
\label{EPAIR1}
&& F_{c}^{\dagger}(\mathbf{r}) F_{v}(\mathbf{r}) = N_c N_v 
\exp{(-\alpha_x x^2 -\alpha_y y^2 - \alpha_z z^2)}, \nonumber \\
&& N_c = \left(\frac{2 \alpha_{xc}}{\pi} \right)^{1/4} 
\left(\frac{2 \alpha_{yc}}{\pi} \right)^{1/4}
\left(\frac{2 \alpha_{zc}}{\pi} \right)^{1/4}, \quad 
N_v = \left(\frac{2 \alpha_{xv}}{\pi} \right)^{1/4} 
\left(\frac{2 \alpha_{yv}}{\pi} \right)^{1/4}
\left(\frac{2 \alpha_{zv}}{\pi} \right)^{1/4}, \nonumber \\
&& \alpha_x = \alpha_{xc} + \alpha_{xv}, \quad 
\alpha_y = \alpha_{yc} + \alpha_{yv}, \quad 
\alpha_z = \alpha_{zc} + \alpha_{zv}, \nonumber \\
&& \alpha_{xc} = \frac{\omega_{x,ee}}{2}, \quad \alpha_{yc} = \frac{\omega_{y,ee}}{2}, 
\alpha_{zc} = \frac{\omega_{z,ee}}{2}, \nonumber \\
&& \alpha_{xv} = \frac{M_{||,hh} \omega_{x,hh}}{2}, 
\quad \alpha_{yv} = \frac{M_{||,hh} \omega_{y,hh}}{2}, 
\alpha_{zv} = \frac{M_{\perp,hh} \omega_{z,hh}}{2}, \nonumber \\
\end{eqnarray}
where $N_c$ and $N_v$ are normalization constants, $\alpha_{\gamma c}$ and 
$\alpha_{\gamma v}$ ($\gamma = x, y, z$) denote confinements of electrons and holes in 
$x$, $y$, and $z$ direction, respectively, and $\alpha_{\gamma} = \alpha_{\gamma c}+
\alpha_{\gamma v}$ denote confinements of electron-hole envelope. 

The matrix elements responsible for ``bright" exciton splitting for $s_e s_h$ exciton 
are given by 
\begin{eqnarray}
&& \delta_{12}^{LRE,N} = \mu^2 \left( R_{xx} - 2i R_{xy} - R_{yy} \right), \nonumber \\
&& \delta_{21}^{LRE,N} = \mu^2 \left( R_{xx} + 2i R_{xy} - R_{yy} \right), \nonumber \\
\end{eqnarray}
where 
\begin{eqnarray}
\label{LRESS}
&& R_{xx} = \int \int \left( \frac{\partial^2 F_{c}^{\dagger} F_{v}}{ \partial x_1^2 } \right)
\frac{F_{c}(\mathbf{r}_2) F_{v}^{\dagger}(\mathbf{r}_2)}{|\mathbf{r}_1-\mathbf{r}_2|}
 d \mathbf{r}_1 d \mathbf{r}_2 = -\frac{(N_c N_v)^2 }{\alpha_x \alpha_y \alpha_z} \pi \sqrt{\pi}
 I_x(\alpha_x,\alpha_y,\alpha_z),  \nonumber \\
&& I_x(\alpha_x,\alpha_y,\alpha_z) = 
 \int_0^{\pi/2} \int_0^{\pi/2}  
 \frac{\sin^3{\theta} \cos^2{\phi} d \theta d \phi}{\left(
 \sin^2{\theta} \left(\frac{1}{2\alpha_x}\cos^2{\phi}+ \frac{1}{2\alpha_y}\sin^2{\phi} \right)+\frac{1}{2\alpha_z}\cos^2{\theta} \right)^{3/2}}\nonumber\\
 && R_{yy} = \int \int \left( \frac{\partial^2 F_{c}^{\dagger} F_{v}}{ \partial y_1^2 } 
 \right)
\frac{F_{c}(\mathbf{r}_2) F_{v}^{\dagger}(\mathbf{r}_2)}{|\mathbf{r}_1-\mathbf{r}_2|}
 d \mathbf{r}_1 d \mathbf{r}_2 = -\frac{(N_c N_v)^2 }{\alpha_x \alpha_y \alpha_z} \pi \sqrt{\pi}
 I_y(\alpha_x,\alpha_y,\alpha_z),  \nonumber \\
&& I_y(\alpha_x,\alpha_y,\alpha_z) = 
 \int_0^{\pi/2} \int_0^{\pi/2}  
 \frac{\sin^3{\theta} \sin^2{\phi} d \theta d \phi}{\left(
 \sin^2{\theta} \left(\frac{1}{2\alpha_x}\cos^2{\phi}+ \frac{1}{2\alpha_y}\sin^2{\phi} \right)+\frac{1}{2\alpha_z}\cos^2{\theta} \right)^{3/2}},  \nonumber\\
 && R_{xy} = \int \int \left(\frac{\partial^2 F_{c}^{\dagger} F_{v}}{ \partial y_1 \partial x_1} \right)
 \frac{F_{c}(\mathbf{r}_2) F_{v}^{\dagger}(\mathbf{r}_2)}{|\mathbf{r}_1-\mathbf{r}_2|}
 d \mathbf{r}_1 d \mathbf{r}_2 = 0 . 
\end{eqnarray}

It follows from~(\ref{LRESS}) that $\delta_{12}^{LRE,N}$ and $\delta_{21}^{LRE,N}$
vanish provided that the ``confinement" of electron-hole pair envelope 
is identical in $x$ and $y$ directions ($\alpha_x = \alpha_y$). Figure 2 illustrates this. 
Figure 2 shows ``bright" exciton doublet splitting as a function of lateral anisotropy $t$. 
The doublet splitting energy is defined as $E_X^{y}-E_X^{x}$, where $E_X^{x}$ and $E_X^{y}$
are ground state ``bright" exciton energy levels polarized along $x$ and $y$ directions, 
respectively. For $t < 0$ ($\omega_x < \omega_y$), $E_X^{y} > E_X^{x}$ and the ``bright" ground state is 
dipole active along the $x$-axis, whereas for $t > 0$ ($\omega_x < \omega_y$) the ``bright" ground state is dipole 
active along the $y$-axis.
In our model, the ``bright" ground state is always dipole-active along the axis of weaker 
confinement. The insert to Figure~2 shows $R_{xx}$ and  $R_{yy}$ ``nonlocal" contributions 
to LRE exchange as a function 
of anisotropy computed from equations~(\ref{LRESS}). 
We can see that $R_{xx}$ increases and $R_{yy}$ decreases 
in magnitude as $t$ goes from -0.5 to 0.5. The insert also shows the difference $R_{xx}-R_{yy}$
which determines ``bright" exciton splitting as a function of anisotropy. We can 
see that $R_{xx}-R_{yy}$ steadily decreases as a function of $t$ and passes through zero at $t=0$.
$t=0$ corresponds to the laterally isotropic confining potential.  In this case, 
$R_{yy} = R_{xx}$ and the exchange matrix element which couple two ``bright" exciton levels
vanishes. As a results, the splitting between ``bright" excitons vanishes as well. 

\subsection{Application of lateral electric field} 

In the previous section, we have demonstrated that the magnitude of the 
``bright" exciton exchange splitting can be controlled through the shape of the 
confining potential. In practice, one may, for example, try to  quench 
the ``bright" exciton splitting by picking ``symmetric" dots from a large number 
of samples grown under different conditions~\cite{Stevenson2006}. Nethertheless, 
the growth of QD is essentially a random process and control through 
the QD shape is hard to achieve. 

It is, therefore, of great interest to examine 
the effects of external fields on the excitonic fine structure. For example, QDs can be placed~\cite{Dalacu2009} between
Schottky gates for the application of vertical and lateral electric fields. 
Recently, it has been demonstrated theoretically~\cite{Korkusinski2009} 
and experimentally~\cite{Reimer2008} that by applying an 
in-plane electric field it is possible to fine-tune photon cascades originating 
from recombination of multiexciton complexes in QDs. 

The electron-hole exchange interaction
was treated in~Ref.~\cite{Korkusinski2009} using empirical exchange Hamiltonian. 
This subsection discusses the effects of the lateral electric field on the exchange matrix 
elements of our effective EMA Hamiltonian~(\ref{MASTER}). We will focus on the 
``bright" exciton splitting as a function of the lateral electric field. 

Within our model, application of lateral electric field 
$\vec{F} = F \mathbf{e}_x + F \mathbf{e}_y$ of magnitude $|F|$ displaces the ``origin" of electron 
and hole envelops in $xy$-plane from $(0,0)$ to $(x_0^e,y_0^e)$ and 
$(x_0^h,y_0^h)$, respectively. The single-particle energy levels are rigidly shifted
by the  Stark shift whereas the spacing between the single-particle energy 
levels of HO is not affected. The ``separation" of electron and hole 
envelopes in $xy$-plane is determined by 
equation
\begin{eqnarray}
&& \Delta x_0 = x_0^e - x_0^h = eF \left( \frac{1}{\omega_{x,ee}^2}+
\frac{1}{M_{||,hh} \omega_{x,hh}^2}  \right) , \nonumber \\
&& \Delta y_0 = y_0^e - y_0^h = eF \left( \frac{1}{\omega_{y,ee}^2}+
\frac{1}{M_{||,hh} \omega_{y,hh}^2}  \right).  \nonumber \\
\end{eqnarray}

Evaluating the integrals which control ``bright" exciton splitting 
$R_{xx}(F), R_{yy}(F), R_{xy}(F)$ as a function of field $F$ for $s_e s_h$ electron-hole 
envelope, we obtain
\begin{eqnarray}
&& R_{xx}(F) = R_{xx}(0) 
\exp{\left(-\frac{2 \alpha_{xc} \alpha_{xv}}{\alpha_x} \Delta x_0^2 -\frac{2 \alpha_{yc} \alpha_{yv}}{\alpha_y} \Delta y_0^2 \right)} , 
\nonumber \\
&& R_{yy}(F) = R_{yy}(0) 
\exp{\left(-\frac{2 \alpha_{xc} \alpha_{xv}}{\alpha_x} \Delta x_0^2 -\frac{2 \alpha_{yc} \alpha_{yv}}{\alpha_y} \Delta y_0^2 \right)} , 
\nonumber \\
&& R_{xy}(F) = 0
\end{eqnarray}
where $R_{xx}(0)$ and $R_{yy}(0) $ are integrals~(\ref{LRESS}) in the absence of 
the electric field ($F=0$). Field dependence of $R_{xx}(F)-R_{yy}(F)$ which determines the 
``bright" exciton splitting  is shown 
on Figure~3 for five different initial values of lateral anisotropy. 
At $F=0$, the ``bright" exciton splitting attains maximum value which is determined by 
the initial shape anisotropy of the confining potential. The larger the magnitude of the anisotropy,
the larger is the initial ($F=0$) ``bright" exciton splitting. 
As $F$ increases in magnitude, the  electron and hole 
envelops are pulled out in opposite direction
and the magnitude 
of the splitting is reduced. Since the separation between the envelopes 
depends on $F^2$, the splitting is independent of the sign of the field $F$.

It is interesting to note that the ``rate of quenching" of the splitting  
depends on the initial anisotropy - the stronger the confinement, the larger is 
the drop in the magnitude of the splitting.  One can use lateral electric field to 
produce two identical ``bright" exciton splittings for two dots with different initial 
anisotropies. This happens, for example, at $|F| \approx 0.6$ for two dots with 
initial anisotropies of $t=-0.5$ and $t=-0.4$, respectively. 

\subsection{Scaling of ``bright" exciton splitting}

The question of size-scaling of exchange interactions in nanosystems has received 
a lot of attention, particularly, in connection with the so-called Stokes (``red") shift of 
resonant PL spectra with respect to the absorption edge 
(see, for example, Ref.~\cite{Efros1996}). In this section, we examine 
the size scaling of ``bright" exciton splitting as a function of system's size.

Suppose that $R$ is a characteristic size of a 2D-like system. 
From the normalization condition
\begin{eqnarray}
 \int d \mathbf{r}  |F_c(\mathbf{r})|^2 = 1, \quad 
 \int d \mathbf{r}  |F_v(\mathbf{r})|^2 = 1, 
\end{eqnarray}
envelope functions scale as $1/R^2$. Based on this ``normalization" argument, 
the nonlocal LRE integrals $R_{xx}$,$R_{yy}$, and $R_{xy}$ that determine ``bright" exciton 
splitting scale as  
\begin{equation}
\label{RSCALE1}
\frac{1}{R^2}  \frac{1}{R^2} \frac{1}{R}   \frac{1}{R^2} R^2 R^2 \propto \frac{1}{R^3} . 
\end{equation} 

The dependence~(\ref{RSCALE1}) is expected to hold in 
the strong confinement regime. Of course, one has to keep in mind that the exchange matrix 
element responsible for the ``bright" exciton splitting is a linear combination of 
nonlocal LRE integrals $R_{xx}$, $R_{yy}$, and $R_{xy}$.  Therefore, the actual 
size dependence of ``bright" exciton splitting might be different from $1/R^3$ and 
depend strongly on the ``nature" of the envelop functions involved in the LRE integrals. 

Suppose the HO potential is of the type $V_{HO}(x,y) =  V_0 ( x^2+y^2)/R^2$, where 
$V_0$ is some characteristic energy scale and $R$ is characteristic size  
determined by HO frequency $\omega^0$, mass $M_{||}$, and characteristic energy scale 
$V_0$
\begin{equation}
R = \frac{1}{\omega^0} \sqrt{\frac{2 V_0}{M_{||}}} = l^2 / L, 
\quad l = \sqrt{ \frac{2}{M_{||} \omega^0}}, \quad L = \sqrt{ \frac{2}{M_{||} V_0}}, 
\end{equation}
where $l$ and $L$ are ``confinement" lengths. If $V_0$ is kept constant,
$ R \propto 1/(M_{||}\omega^0)$. We will, therefore, calculate the ``bright" exciton splitting 
as a function of $1/(M_{||} \omega^0)$ keeping the ratio of electron and hole confinement 
frequencies constant $\omega_{ee}^0 / \omega_{hh}^0 = const$ and the lateral anisotropy $t$
fixed. Since $\omega_{ee}^0 / \omega_{hh}^0 = const$ by increasing/decreasing 
$\omega_{ee}^0$ we automatically increasing/decreasing $\omega_{hh}^0$. 
 
Figure 4 shows the  scaling of the splittings for the ground and excited ``bright" exciton levels with size. 
The lateral anisotropy  is fixed to $t=-0.1$ in all the calculations. We find that splittings decay as 
the size of the system increases. We find that the ground state exciton 
splitting is size-insensitive whereas the splittings of the excited bright excitons 
scale $\propto 1/R^{1.3}$. This scaling is different from $1/R^3$ dependence expected from 
the ``normalization" arguments~(\ref{RSCALE1}). The deviation might be due to the 
fact that the exchange 
matrix elements that determine the ``bright" exciton splitting $\delta_{12}$ and $\delta_{21}$ 
involve a linear combination of nonlocal exchange integrals $R_{xx}$, $R_{yy}$, and $R_{xy}$. 
Moreover, $R_{xx}$ and $R_{yy}$ enter the expression for $\delta_{12}$ and $\delta_{21}$
with different sign and some cancellation of terms may occur. Therefore,  
the size dependence of ``bright" exciton splittings may depend strongly on the details 
of the excitonic wavefunction. 

\section{Conclusions}

A four-band electron-hole excitonic Hamiltonian is derived  
within the effective mass approximation (EMA) which takes into account the electron-hole 
exchange interaction. The matrix elements of the 
effective Hamiltonian are expressed explicitly 
in terms of electron and hole envelopes. The ``microscopic" parts of single-particle orbitals 
are integrated out and enter our Hamiltonian implicitly through numerical parameters. 
The matrix element  responsible for the ``bright" exciton splitting is identified and 
analyzed. An explicit expression for this matrix element in terms of electron and hole 
envelopes is presented. An excitonic fine structure for a model system with 2D-like anisotropic 
HO confining potential is considered. It is found that in the case of isotropic potential, 
the ``bright" exciton splitting vanishes. Within  the formalism of our effective excitonic Hamiltonian, 
the effects of the lateral electric field on the excitonic fine structure were considered. It is 
found that the excitonic structure can be tuned with lateral electric field and that the magnitude 
of the exchange splitting is reduced but never to zero. The origin of this reduction is the separation of electron-hole pairs. 
Finally, size-dependence of the ``bright" exciton splittings for the ground and excited state excitons 
was investigated and it was found that the size scaling of ``bright" exciton splitting can be different 
from the laws established using normalization conditions.  

\section{Acknowledgment}
The authors acknowledge discussions with M. Korkusinki, M. Zielinski, R. Williams and D. Dalacu 
and support by the NRC-NSERC-BDC Nanotechnology  project, QuantumWorks, NRC-CNRS CRP and CIFAR. 
The authors thank Prof. E. L. Ivchenko for reading the manuscript and making useful suggestions. 
\appendix

\bibliographystyle{apsrev}

\newpage

\begin{center}
\begin{figure*}
\begin{minipage}{1.0\textwidth}
\caption{Single-particle and Exciton Energy Levels as a function of lateral anisotropy $t$ 
of the confining potential.}
\end{minipage}
\begin{center}
\begin{minipage}{1.0\textwidth}
\subfigure[~]
{\epsfig{file=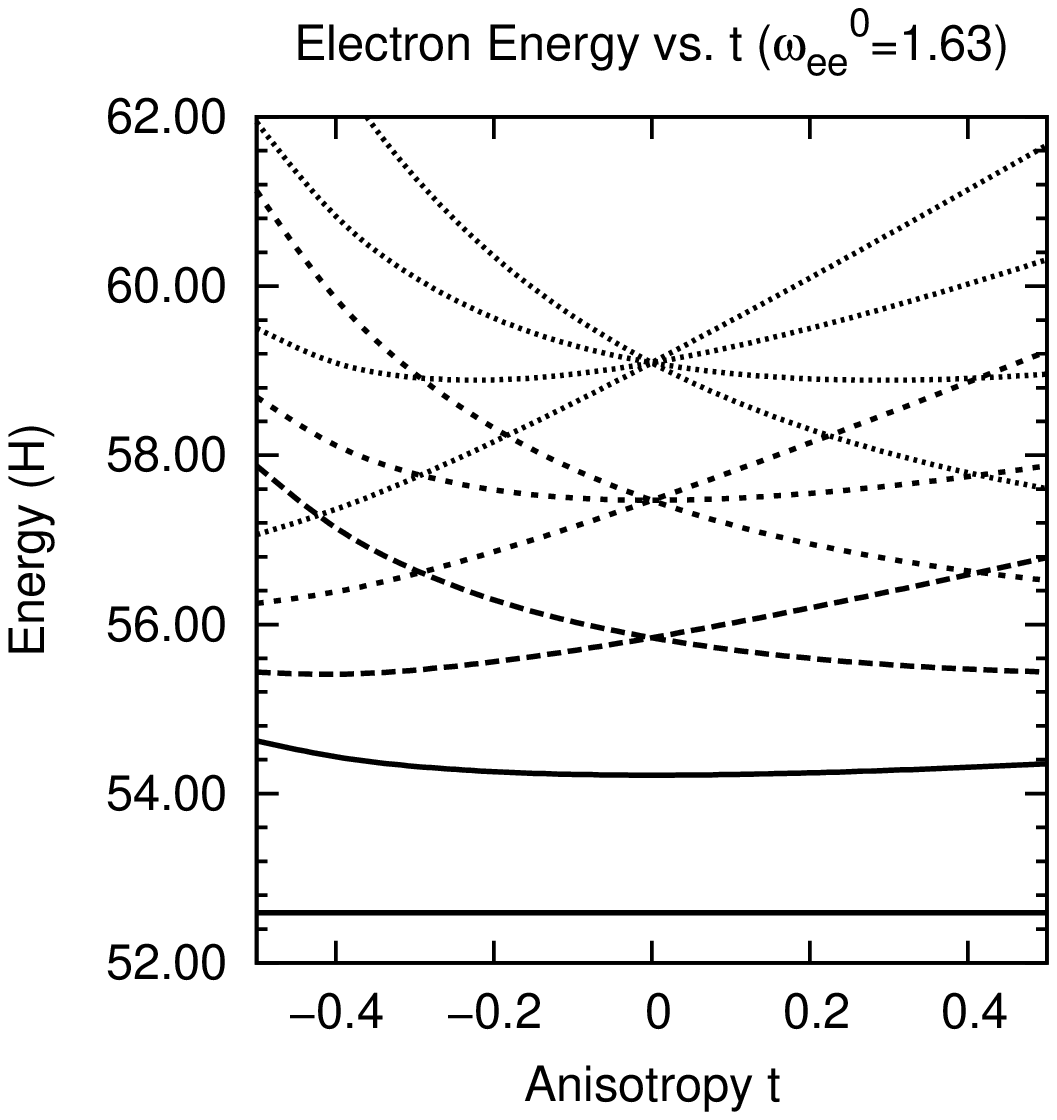,
width=0.40\textwidth,keepaspectratio,angle=270}}
\hfill
\subfigure[~]
{\epsfig{file=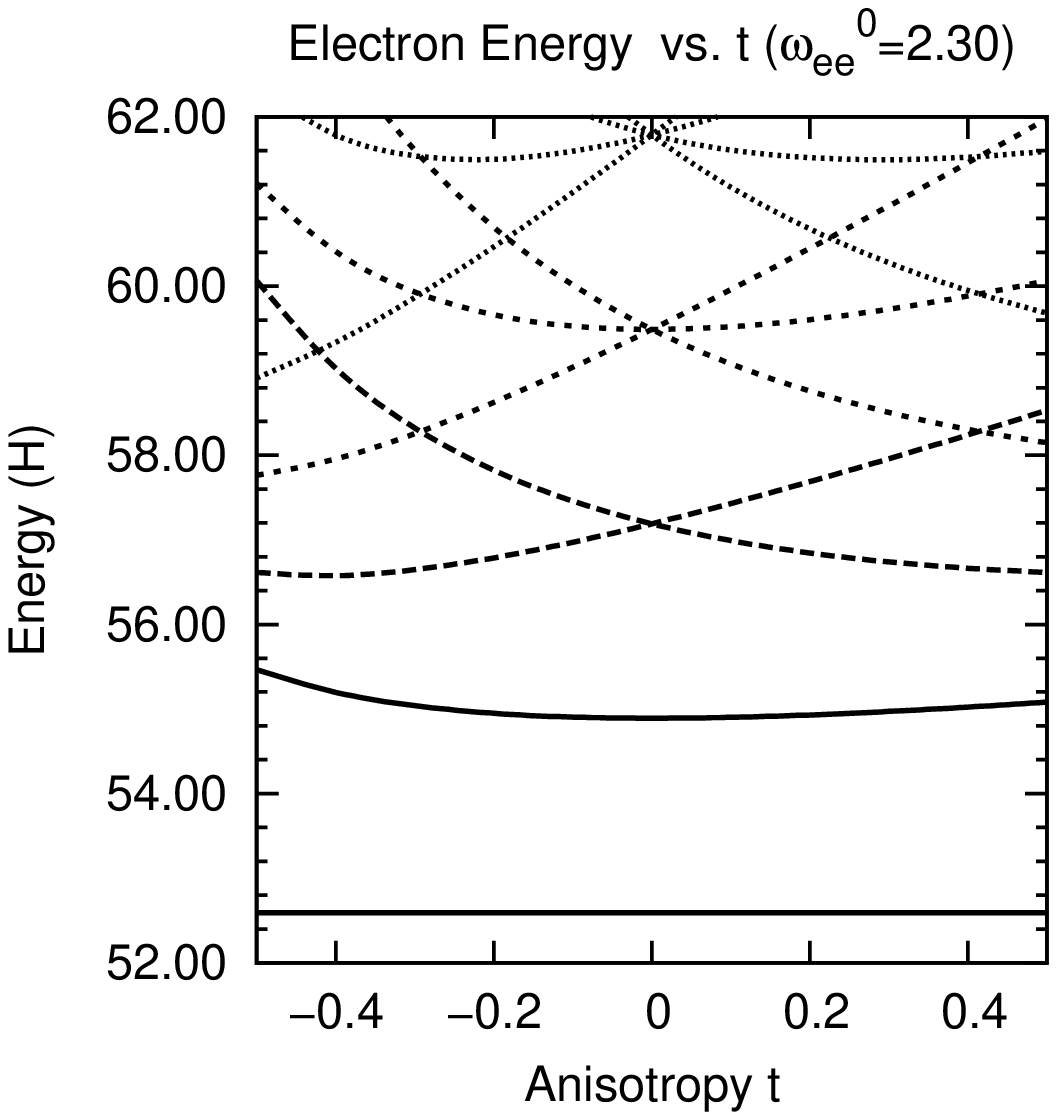,
width=0.40\textwidth, keepaspectratio,angle=270}}
\\
 \subfigure[~]
{\epsfig{file=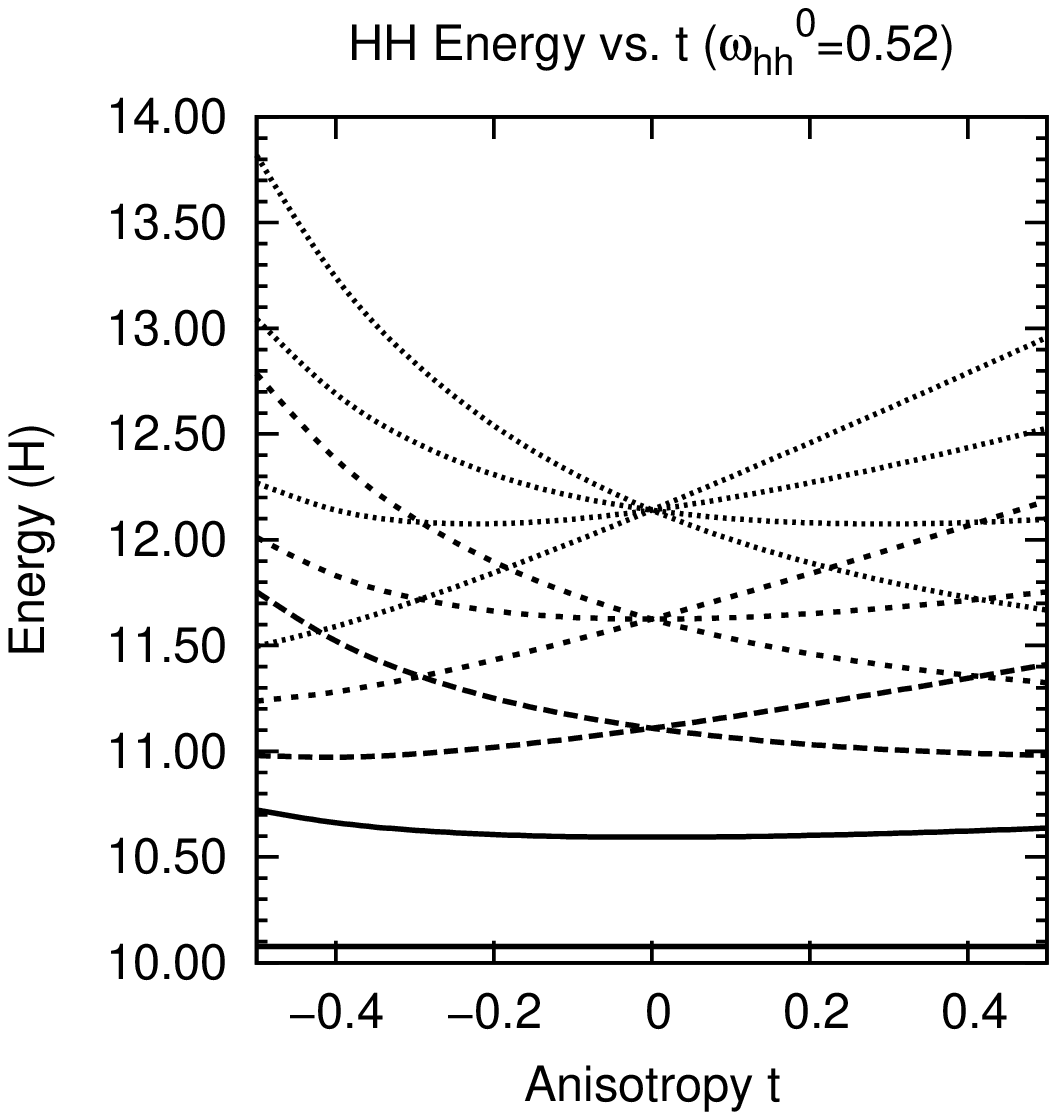,
width=0.40\textwidth, keepaspectratio,angle=270}}
\hfill
\subfigure[~]
{\epsfig{file=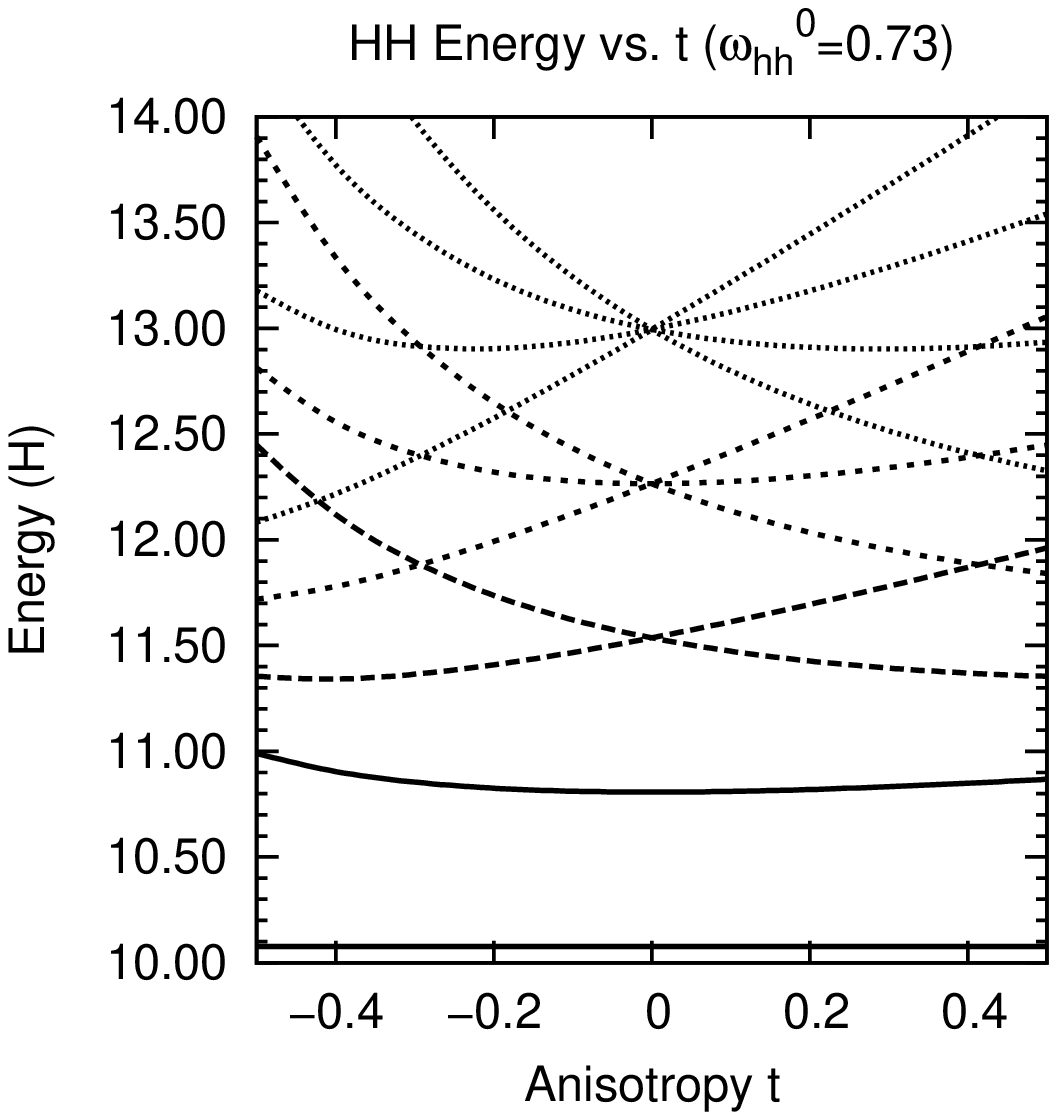,
width=0.40\textwidth, keepaspectratio,angle=270}}
\\
 \subfigure[~]
{\epsfig{file=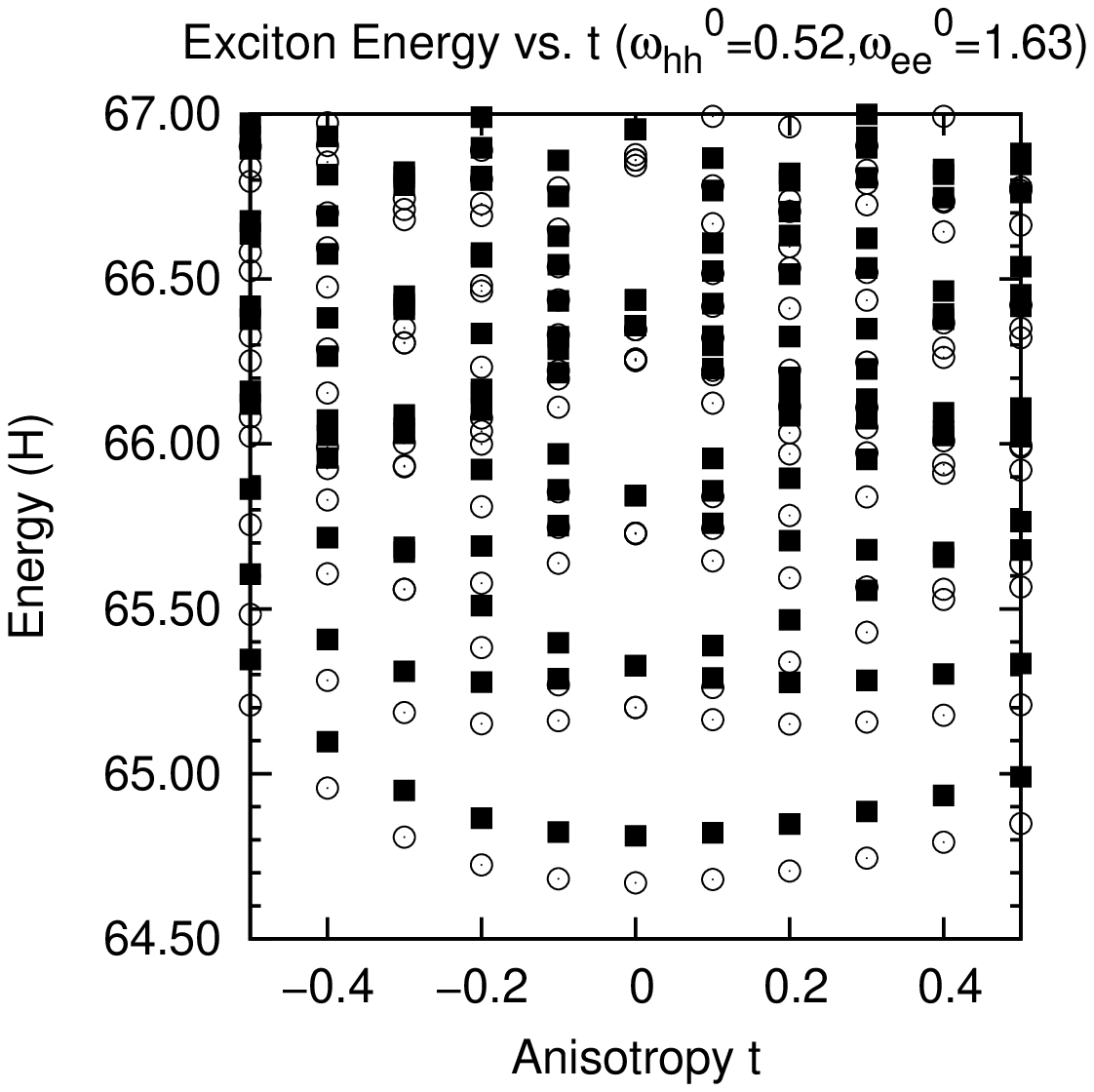,
width=0.40\textwidth, keepaspectratio,angle=270}}
\hfill
\subfigure[~]
{\epsfig{file=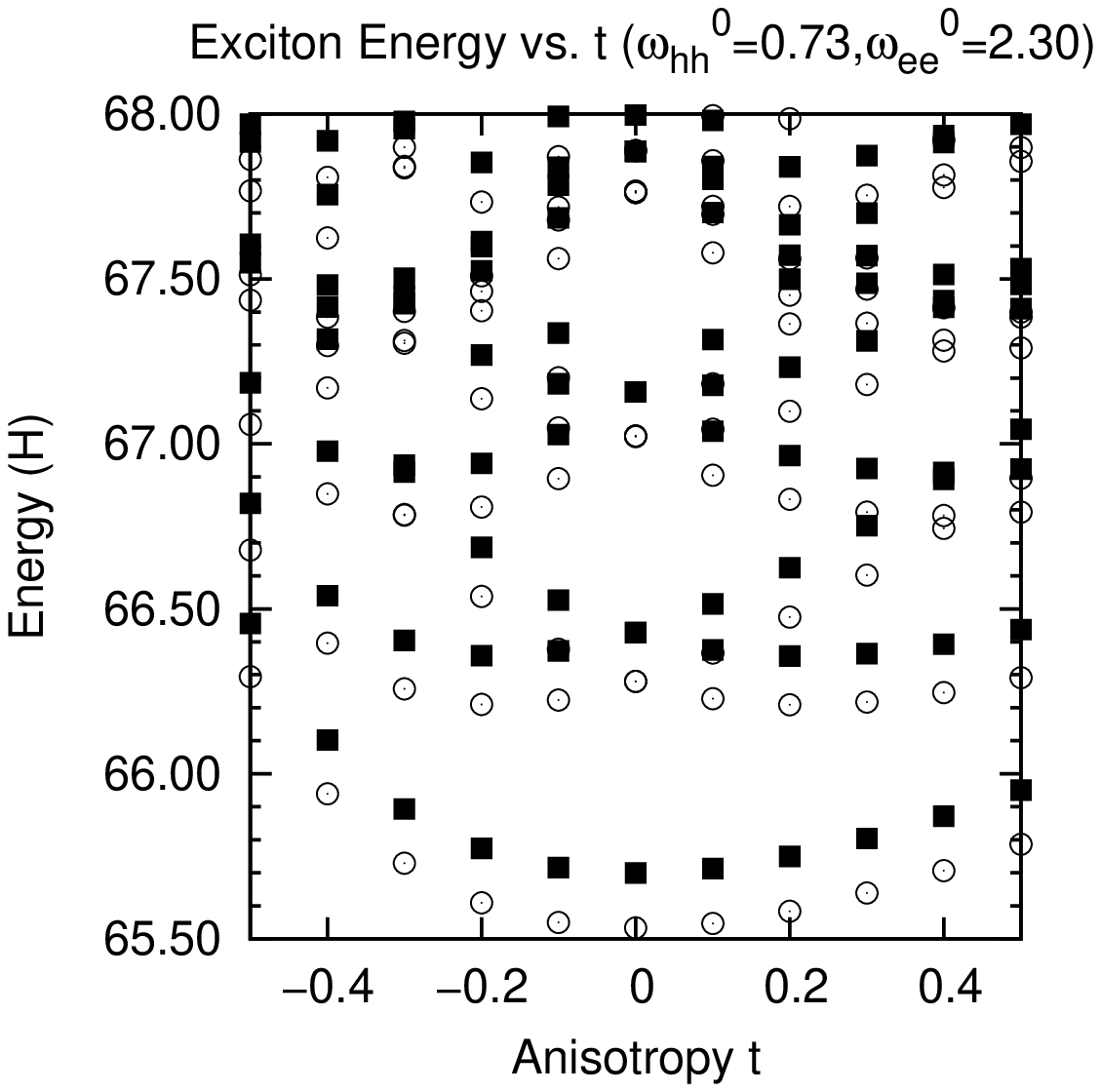,
width=0.40\textwidth, keepaspectratio,angle=270}}
\end{minipage}
\end{center}
\end{figure*}
\end{center}

\begin{center}
\begin{figure*}
\begin{minipage}{1.0\textwidth}
\caption{``Bright" exciton doublet splitting as a function of lateral anisotropy $t$
of the confining potential. Insert: $R_{xx}$ (squares),  
$R_{yy}$ (circles) contributions to ``nonlocal" LRE matrix elements 
as a function of anisotropy of the confining potential;  
$R_{xx}-R_{yy}$ (stars) difference of two ``nonlocal" LRE contributions 
that determines the ``bright" exciton splitting. For the isotropic 
confining potential $t=0$ and $R_{xx}-R_{yy}=0$.}
\end{minipage}
\begin{center}
\begin{minipage}{1.0\textwidth}
\epsfig{file=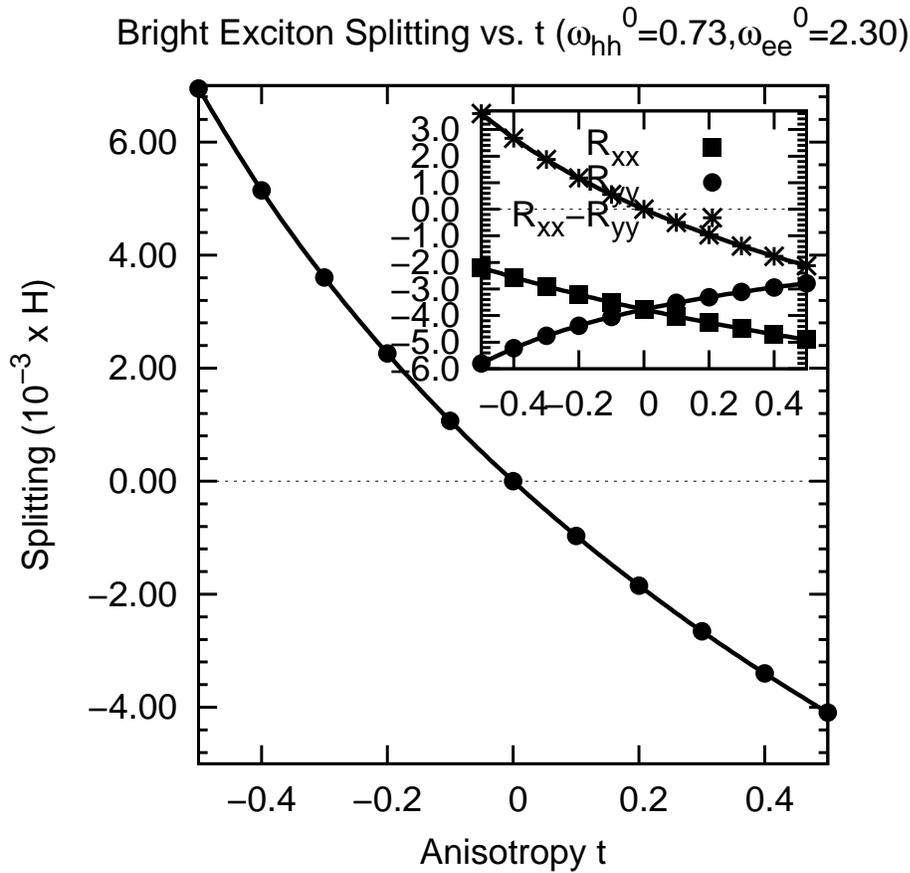,width=1.2\textwidth,keepaspectratio,angle=270}
\end{minipage}
\end{center}
\end{figure*}
\end{center}

\begin{center}
\begin{figure*}
\begin{minipage}{1.0\textwidth}
\caption{``Nonlocal" LRE $R_{xx}-R_{yy}$ 
which determines ``bright" exciton splitting as function of lateral electric field $F$. 
At $F=0$, $R_{xx}-R_{yy}$ attains maximum determined by the initial anisotropy of the 
confining potential. As magnitude of field 
$F$ increases, $R_{xx}-R_{yy}$ decreases due to the separation of electron and hole 
envelopes.}
\end{minipage}
\begin{center}
\begin{minipage}{1.0\textwidth}
\epsfig{file=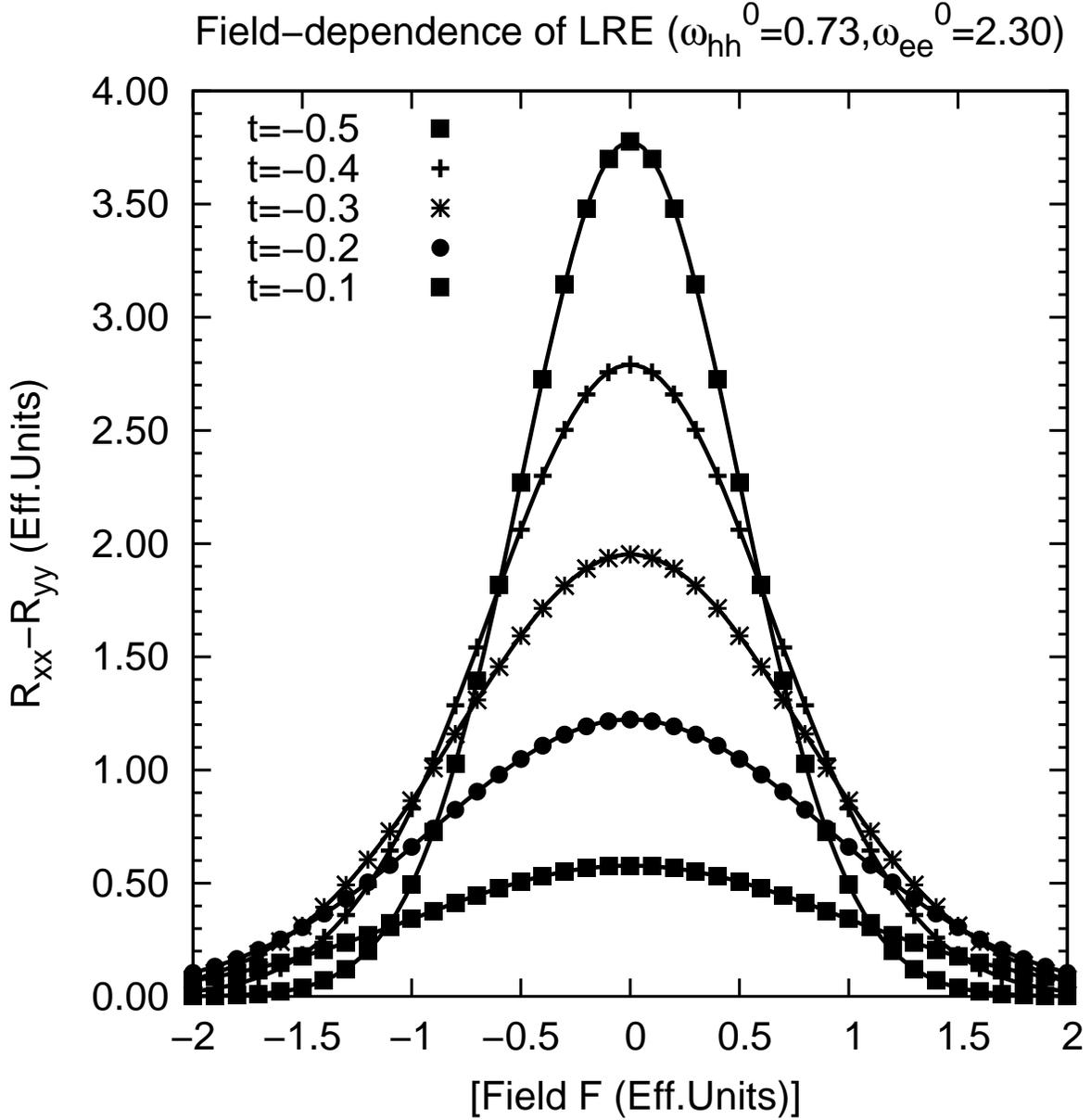,width=1.0\textwidth,keepaspectratio,angle=270}
\end{minipage}
\end{center}
\end{figure*}
\end{center}

\begin{center}
\begin{figure*}
\begin{minipage}{1.0\textwidth}
\caption{Size scaling of the ``bright" exciton doublet splittings. Lateral
anisotropy is kept constant $t=-0.1$. The doublet splittings are plotted as a 
function  of $1/(M_{||,ee}\omega_{ee}^0)$.  The ratio $\omega_{ee}^0/\omega_{hh}^0$ is kept constant. 
The ground exciton is  of $s_e-s_h$ type. Two
excited ``bright"  exitons correspond to  $s_e-d_h$ electron-hole pairs. The solid lines 
are fit to the power law $C/R^n$, where $C$ is a constant, $n \approx 1.30$ .  }
\end{minipage}
\begin{center}
\begin{minipage}{1.0\textwidth}
\epsfig{file=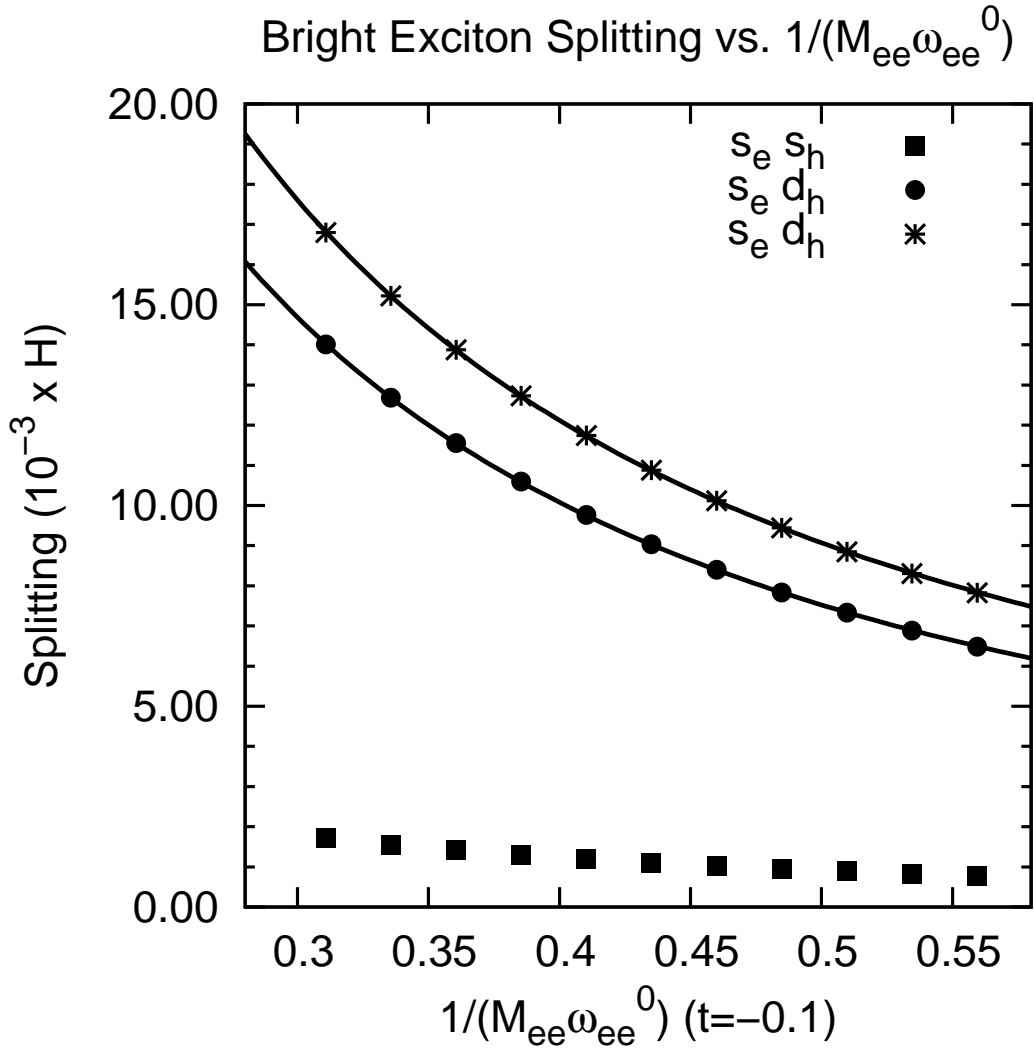,width=0.9\textwidth,keepaspectratio,angle=270}
\end{minipage}
\end{center}
\end{figure*}
\end{center}

\end{document}